\documentclass[11pt,a4paper]{article}

\usepackage[utf8]{inputenc}
\usepackage[T1]{fontenc}
\usepackage{lmodern}
\usepackage{amsmath,amssymb}
\usepackage{graphicx}
\usepackage{booktabs}
\usepackage{hyperref}
\usepackage{xcolor}
\usepackage{listings}
\usepackage{multirow}
\usepackage{url}
\usepackage{tabularx}
\usepackage{authblk}
\usepackage[margin=1in]{geometry}

\hypersetup{
  colorlinks=true,
  linkcolor=blue!70!black,
  citecolor=green!50!black,
  urlcolor=blue!70!black
}

\begin{document}

\title{Security Is Relative: Training-Free Vulnerability Detection\\via Multi-Agent Behavioral Contract Synthesis}

\author[1]{Yongchao Wang}
\author[1]{Zhiqiu Huang}
\affil[1]{College of Computer Science and Technology, Nanjing University of Aeronautics and Astronautics, Nanjing, China}
\affil[ ]{\texttt{\{wangyc, zqhuang\}@nuaa.edu.cn}}

\date{April 2026}

\maketitle

\begin{abstract}
Automated vulnerability detection using deep learning has shown promising results on early benchmarks, but recent rigorous evaluations reveal catastrophic performance degradation: models that achieved F1~$>$~0.68 on legacy datasets collapse to as low as 0.031 under strict deduplication and temporal splitting.
We identify the root cause as the \textit{semantic ambiguity problem}: identical code patterns can be secure or vulnerable depending on project-specific behavioral contracts, rendering any global classification approach fundamentally inadequate.
We propose \textbf{Phoenix}, a training-free multi-agent framework that resolves this ambiguity through \textit{Behavioral Contract Synthesis}.
Phoenix decomposes vulnerability detection into three specialized stages: a \textit{Semantic Slicer} that extracts minimal vulnerability-relevant context, a \textit{Requirement Reverse Engineer} that synthesizes Gherkin behavioral specifications encoding the security contract distinguishing vulnerable from patched code, and a \textit{Contract Judge} that evaluates code against these specifications as a strict compliance check.
On the PrimeVul Paired Test Set, Phoenix achieves \textbf{F1~=~0.825} and \textbf{Pair-Correct~=~64.4\%}, substantially surpassing the prior state-of-the-art RASM-Vul (F1~=~0.668, P-C~=~21.4\%) and VulTrial (F1~=~0.563, P-C~=~18.6\%) while using fully open-source models that are up to 48$\times$ smaller (7--14B vs.\ DeepSeek-V3 671B).
Comprehensive ablation across 25 configurations and 5 model families demonstrates that Gherkin specifications are the decisive performance driver (+0.09 to +0.35 F1).
Qualitative error analysis reveals that 18\% (17/97) of Phoenix's ``False Positives'' identify genuine security concerns in developer-patched code, empirically demonstrating that security is not an absolute property of code syntax, but a relative property defined against specific behavioral contracts.
\end{abstract}


\section{Introduction}
\label{sec:introduction}

Automated software vulnerability detection is among the most critical challenges in modern software engineering.
In 2025, over 48,000 Common Vulnerabilities and Exposures (CVEs) were published, a 20\% increase from approximately 40,000 in 2024~\cite{cvemetrics}.
This escalating threat landscape, combined with the accelerating integration of generative AI into the software development lifecycle, has created an urgent need for detection systems that can match the pace and scale of modern software production.

Deep learning approaches to vulnerability detection --- including token-sequence models (VulDeePecker~\cite{vuldeepecker}, SySeVR~\cite{sysevr}), graph neural networks (Devign~\cite{devign}, ReVeal~\cite{reveal}), and pre-trained code models (CodeBERT~\cite{codebert}, UniXcoder~\cite{unixcoder}) --- have demonstrated promising results on early benchmarks.
However, the introduction of the \textbf{PrimeVul} benchmark (ICSE 2025)~\cite{primevul} exposed a sobering reality: when evaluated under strict deduplication, temporal splitting, and precise labeling protocols, models that achieved F1 scores above 0.68 on the legacy Big-Vul dataset collapsed to as low as 0.031 on PrimeVul.
Even advanced 7-billion-parameter models such as StarCoder2, which scored 68.26\% F1 on Big-Vul, plummeted to a mere 3.09\% --- indistinguishable from random guessing~\cite{primevul}.
GPT-3.5 and GPT-4 with supervised fine-tuning similarly failed under these rigorous conditions~\cite{primevul}.
These findings reveal that prior models were not learning genuine vulnerability semantics but merely overfitting to superficial distributional artifacts within flawed datasets.

The root cause of this failure is what we term the \textbf{semantic ambiguity problem}: identical or structurally similar code patterns can be secure in one project and critically vulnerable in another, depending on project-specific sanitizers, architectural conventions, and runtime contexts~\cite{cprvul}.
A function that passes raw user input to a database query is vulnerable to SQL injection in isolation, but is perfectly secure if the project's middleware architecture guarantees pre-sanitization.
This \textit{contextual dependency} --- with recent analysis showing that over 90\% of real-world vulnerabilities depend on context beyond the function boundary~\cite{crossrepo} --- renders any ``global standard'' classification approach fundamentally inadequate.

This ambiguity is not merely theoretical.
Our analysis of the PrimeVul dataset reveals concrete instances of what we call the \textbf{Double Standard Problem}: within the same project, code with \textgreater 99\% sequence similarity (measured via character-level sequence matching) receives opposite vulnerability labels depending on which CVE it is evaluated against.
For example, in the \texttt{mruby} project, the function \texttt{gen\_assignment()} in \texttt{codegen.c} appears with 100\% code similarity across two entries --- one labeled ``good'' (the CVE-2022-0717 fix) and one labeled ``bad'' (still vulnerable to CVE-2022-1276) --- despite sharing the same CWE category (CWE-125, Out-of-bounds Read).
Similarly, TensorFlow's \texttt{FractionalAvgPoolGrad} function in \texttt{fractional\_avg\_pool\_op.cc} appears at 99.9\% character-level similarity with opposite labels across CVE-2021-37651 and CVE-2022-21730.
We identified 10 such cross-CVE pairs with $>$75\% code similarity in the PrimeVul Paired Test Set alone (see Appendix~\ref{app:casestudies}).
These findings empirically demonstrate that \textit{security is not an absolute property of code syntax, but a relative property defined against specific behavioral contracts} --- precisely the property that Phoenix's Gherkin-based approach is designed to capture.

Recent research has pursued three principal directions to address this challenge, each encountering distinct limitations:

\begin{enumerate}
    \item \textbf{Retrieval-Augmented Generation (RAG):} Systems such as RASM-Vul~\cite{rasmvul} augment LLMs with multi-view retrieval over historical vulnerability databases, achieving F1~=~0.668 on PrimeVul Paired using DeepSeek-V3 (671B parameters). However, these approaches remain fundamentally anchored to \textit{pattern similarity} rather than behavioral intent, struggling with zero-day logic flaws that have no historical analogues.

    \item \textbf{Agentic Static Analysis Synthesis:} QLCoder (ICLR 2026)~\cite{qlcoder} employs LLM agents to synthesize CodeQL queries from vulnerability-fix pairs, achieving F1~=~0.70 on its target Java benchmark. While innovative, it is constrained by the fragility of domain-specific query languages, heavy compilation dependencies, and language-specific toolchains.

    \item \textbf{Context-Enriched Reasoning:} VulnAgent-X~\cite{vulnagentx} introduces a staged auditing pipeline with threshold-based escalation from screening to deep analysis. CPRVul~\cite{cprvul} couples context profiling with inter-procedural reasoning, but finds that naively appending context often degrades performance due to attention dilution --- additional context becomes noise rather than signal when models lack explicit guidance on what to verify.
\end{enumerate}

In this paper, we propose \textbf{Phoenix}, a training-free multi-agent framework that resolves the semantic ambiguity problem through \textit{Behavioral Contract Synthesis}.
Rather than asking an LLM to classify code based on learned patterns or retrieved examples, Phoenix decomposes vulnerability detection into three specialized stages:
(1)~a \textbf{Semantic Slicer} that extracts the minimal vulnerability-relevant code context,
(2)~a \textbf{Requirement Reverse Engineer} that synthesizes a Gherkin specification encoding the precise security contract distinguishing vulnerable code from its patch, and
(3)~a \textbf{Contract Judge} that evaluates individual code samples against this specification as a strict compliance check.

The key innovation is the introduction of \textbf{Gherkin specifications} --- structured behavioral contracts expressed in human-readable Given-When-Then format --- as an explicit intermediate representation between code analysis and classification.
By externalizing the security contract, Phoenix transforms vulnerability detection from an open-ended pattern-matching task into a closed-form contract verification problem.
This decomposition provides project-specific, sample-specific decision criteria to the judge, resolving the contextual ambiguity that defeats monolithic approaches.

\textbf{Contributions.} Our contributions are as follows:
\begin{enumerate}
    \item We propose Phoenix, a training-free, zero-shot multi-agent architecture for vulnerability detection that achieves \textbf{F1~=~0.825} and \textbf{Pair-Correct~=~64.4\%} on the PrimeVul Paired Test Set, substantially surpassing the prior state-of-the-art RASM-Vul (F1~=~0.668) and VulTrial (F1~=~0.563) while using fully open-source models up to 48$\times$ smaller (7--14B vs.\ DeepSeek-V3 671B).

    \item We introduce the novel use of Gherkin behavioral specifications as structured intermediate representations for vulnerability reasoning, bridging the gap between code-level analysis and binary classification through explicit behavioral contracts.

    \item We conduct a comprehensive ablation study across 25 experimental configurations and 5 model families, demonstrating that (a)~Gherkin specifications are the decisive performance driver (+0.09 to +0.35 F1), (b)~specification quality directly correlates with detection performance, and (c)~the approach is model-agnostic, consistently improving performance across diverse LLM architectures.
\end{enumerate}

\section{Related Work}
\label{sec:related}

\subsection{Deep Learning for Vulnerability Detection}

Early deep learning approaches framed vulnerability detection as a supervised binary classification problem over code representations.
VulDeePecker~\cite{vuldeepecker} pioneered the use of code gadgets --- semantically related code slices --- as input to bidirectional LSTMs.
SySeVR~\cite{sysevr} extended this with program slicing based on syntax, semantics, and vector representations.
Devign~\cite{devign} introduced graph neural networks over composite code representations (AST, CFG, DFG), while ReVeal~\cite{reveal} combined graph-based features with gated recurrent units.

The advent of pre-trained code models brought significant progress.
CodeBERT~\cite{codebert}, GraphCodeBERT~\cite{graphcodebert}, and UniXcoder~\cite{unixcoder} leveraged large-scale pre-training on code corpora to produce transferable representations.
LineVul~\cite{linevul} applied these representations to achieve state-of-the-art results on Big-Vul.
However, the PrimeVul benchmark~\cite{primevul} demonstrated that these results were inflated by data leakage and label noise: under rigorous evaluation, all pre-trained models experienced catastrophic performance degradation, with some dropping from F1~$>$~0.60 to below 0.05~\cite{primevul}.

This collapse reveals a fundamental limitation: these models learn distributional surface features (token sequences, syntactic patterns) rather than the semantic reasoning required for genuine vulnerability understanding.
A systematic analysis of ML-based vulnerability detection benchmarks further confirms this, showing that models exploit spurious correlations (e.g., word frequency patterns) rather than learning genuine vulnerability semantics~\cite{crossrepo}.

\subsection{Retrieval-Augmented Approaches}

To address the knowledge limitations of standalone models, retrieval-augmented generation (RAG) systems inject external evidence during inference.
RASM-Vul~\cite{rasmvul} represents the state-of-the-art in this category, constructing a multi-view vector knowledge base with five specialized indices (code semantics, AST structure, vulnerability knowledge, line-change patterns, and AST-change patterns), unified through a Weighted Reciprocal Ranking Fusion (WRRF) algorithm.
Evaluated with DeepSeek-V3 (671B MoE, 37B active parameters) as the backbone, RASM-Vul achieves F1~=~0.668 and Pair-Correct~=~21.4\% on the PrimeVul Paired dataset~\cite{rasmvul}.

Despite this achievement, RAG-based approaches are fundamentally constrained by their reliance on \textit{similarity-based reasoning}.
They answer ``what does this code look like relative to known vulnerabilities?'' rather than ``what should this code do and does it violate that contract?''
This makes them vulnerable to: (a)~zero-day logic flaws with no historical analogues in the knowledge base, (b)~semantically similar but causally opposite code pairs where a single-character fix distinguishes vulnerable from patched versions~\cite{codebert}, and (c)~project-specific business logic vulnerabilities that cannot be captured by general-purpose vector similarity~\cite{reveal}.

Recent work has also explored external knowledge structures to guide LLM reasoning.
VulReaD~\cite{vulread} constructs CWE-level knowledge graphs to support root-cause reasoning, enabling models to leverage structured vulnerability taxonomies during analysis.
However, these approaches fundamentally frame detection as \textit{global classification} --- reasoning about whether code matches known vulnerability categories --- rather than verifying compliance against a code-specific behavioral contract.

\subsection{Multi-Agent Frameworks for Code Security}

The emergence of LLM-based agent systems has introduced more sophisticated approaches to vulnerability analysis.
\textbf{QLCoder} (ICLR 2026)~\cite{qlcoder} uses an agentic framework to synthesize CodeQL queries from CVE metadata and vulnerability-fix pairs.
It introduces an execution feedback loop where generated queries are validated against both vulnerable and patched code versions, achieving a 53.4\% success rate and F1~=~0.70 on a Java CVE benchmark.
However, QLCoder's dependence on the CodeQL infrastructure --- which requires full project compilation, is restricted to supported languages, and suffers from DSL syntax fragility --- limits its practical applicability~\cite{qlcoder}.

\textbf{vEcho}~\cite{vecho} shifts from passive verification to proactive discovery through an Echoic Vulnerability Propagation (EVP) mechanism with a Cognitive Memory Module, enabling the system to learn from confirmed findings and discover analogous vulnerabilities outside the initial rule base.

\textbf{VulnAgent-X}~\cite{vulnagentx} introduces a layered agentic framework with staged evidence fusion, treating vulnerability detection as a progressive auditing process with threshold-based escalation from screening to deep analysis.

\textbf{CPRVul}~\cite{cprvul} couples context profiling with structured reasoning, constructing Code Property Graphs (CPGs) to extract inter-procedural context.
A key finding of CPRVul is that inter-procedural context improves performance only when models are explicitly trained to reason over it --- without structured reasoning guidance, additional context degrades performance by acting as noise~\cite{cprvul}.
This finding directly motivates our approach: rather than training models on context, we provide structured Gherkin specifications as explicit reasoning scaffolding.

Concurrent to our work, \textbf{VulTrial} (ICSE 2026)~\cite{vultrial} proposes a courtroom-inspired multi-agent framework where specialized agents --- a Security Researcher (prosecutor), Code Author (defense), Moderator (judge), and Review Board (jury) --- debate code vulnerabilities through adversarial argumentation.
VulTrial achieves P-C~=~81/435 (18.6\%) on PrimeVul Paired with GPT-4o, demonstrating that multi-perspective reasoning improves over single-agent baselines~\cite{vultrial}.
However, VulTrial's open-ended debate paradigm lacks a formal intermediate representation: agents argue about whether code ``appears'' vulnerable, but without an explicit definition of what constitutes a vulnerability \textit{for the specific code under analysis}.
Phoenix differs fundamentally by introducing Gherkin behavioral contracts as a structured intermediate representation, transforming the open question ``Is this code vulnerable?'' into the closed verification ``Does this code satisfy this specific contract?''
This distinction is reflected in the performance gap: Phoenix achieves P-C~=~64.4\% with 7--14B open-source models, compared to VulTrial's 18.6\% with GPT-4o.

\subsection{Behavioral Specifications and Behavior-Driven Development}

Behavior-Driven Development (BDD) and its specification language Gherkin have been widely adopted in software testing to bridge the gap between stakeholder requirements and executable test cases~\cite{testpyramid}.
Recent work has explored the intersection of BDD with AI-driven development.
AgileGen~\cite{agilegen} uses Gherkin acceptance criteria to align LLM-generated code with user stories, demonstrating that structured behavioral specifications improve semantic consistency in code generation.
Amazon's security team has explored leveraging generative AI for automated Gherkin script generation to formalize AWS security controls~\cite{amazonbdd}.

More importantly, recent empirical studies demonstrate that BDD is particularly effective for extracting and verifying \textit{security} non-functional requirements, with LLM-generated Gherkin scenarios achieving high fidelity in formalizing security controls~\cite{bddsecurity}.
Closer to our domain, VulInstruct~\cite{specguided} proposes teaching LLMs root-cause reasoning by extracting security specifications from vulnerability-fix pairs, and uses these specifications to guide detection.
However, VulInstruct's specifications remain natural-language descriptions that still require open-ended reasoning for classification.

In our own prior work, \textbf{Project Prometheus}~\cite{prometheus} demonstrated the power of reverse-engineered Gherkin specifications in the domain of \textit{Automated Program Repair} (APR).
Prometheus deploys a structurally analogous multi-agent pipeline --- an \textit{Architect} that infers BDD specifications from runtime failures, an \textit{Engineer} that validates them via ``Sandwich Verification,'' and a \textit{Fixer} that performs specification-guided repair --- achieving a 93.97\% fix rate on 680 Defects4J bugs.
The present work extends this paradigm from program repair to vulnerability detection, demonstrating that Gherkin-based behavioral contracts are a \textit{cross-domain} enabling technology: in APR, they guide \textit{what to fix}; in vulnerability detection, they define \textit{what to verify}.

To our knowledge, Phoenix is the first system to automatically reverse-engineer \textit{strict Gherkin behavioral specifications} from vulnerability-fix pairs and enforce them as \textit{rigid, zero-shot compliance verification criteria}, combining the formalism of BDD with the automation of multi-agent LLM reasoning.

\section{Methodology}
\label{sec:methodology}

\begin{figure*}[t]
\centering
\includegraphics[width=\textwidth]{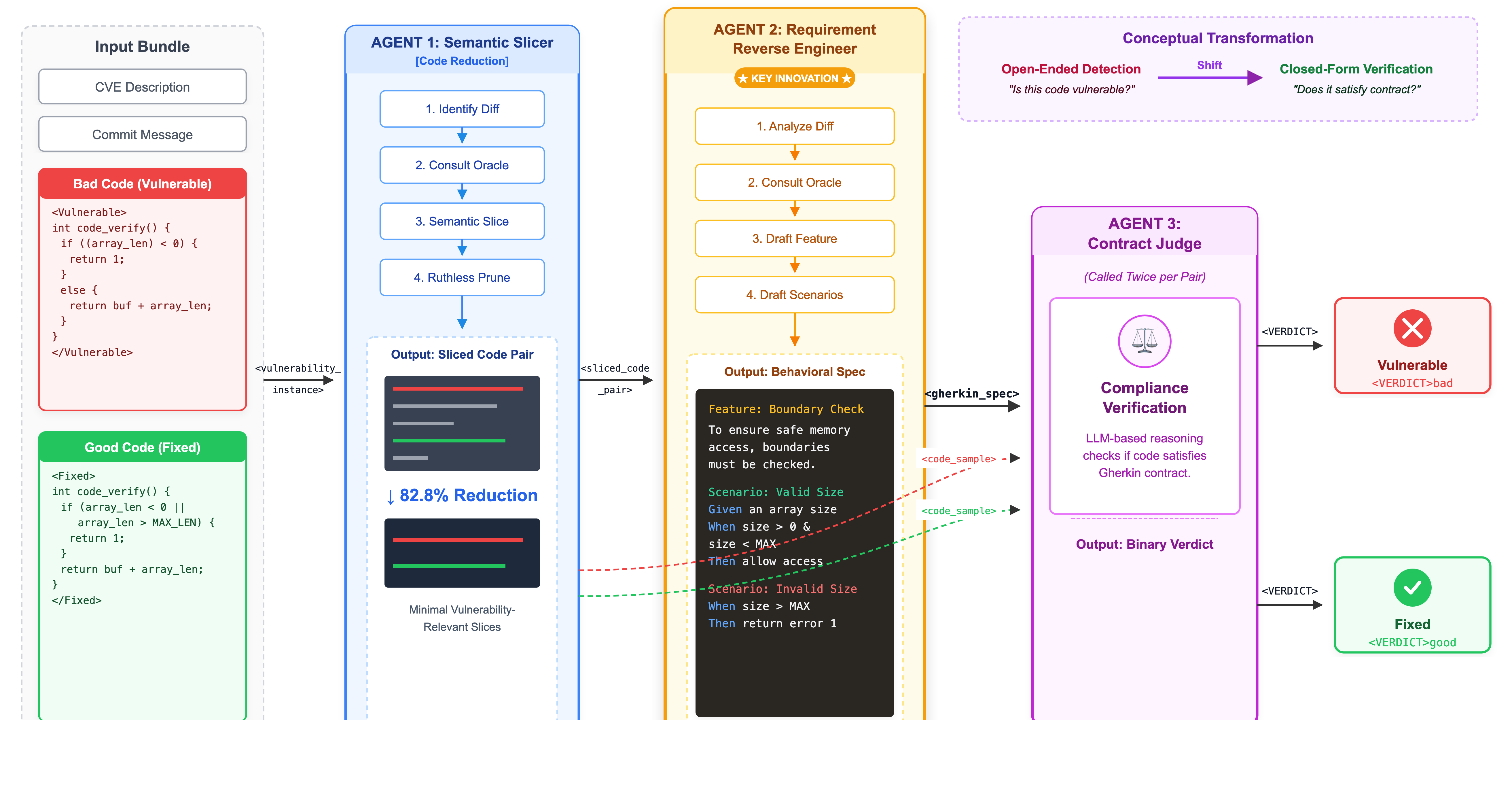}
\caption{Overview of the Phoenix multi-agent framework. The pipeline processes paired code (vulnerable + fixed) through three stages: (1)~Agent~1 (Semantic Slicer) extracts minimal vulnerability-relevant code slices, achieving 82.8\% average code reduction; (2)~Agent~2 (Requirement Reverse Engineer) synthesizes a Gherkin behavioral specification encoding the security contract; (3)~Agent~3 (Contract Judge) evaluates each code sample against the specification, producing a binary verdict. The key innovation is the transformation from open-ended vulnerability detection (``Is this code vulnerable?'') to closed-form contract verification (``Does this code satisfy the behavioral contract?'').}
\label{fig:framework}
\end{figure*}

\subsection{Overview}

We propose \textbf{Phoenix}, a training-free multi-agent framework that decomposes zero-shot vulnerability detection into three specialized inference stages: \textit{Semantic Slicing}, \textit{Requirement Reverse Engineering}, and \textit{Contract-Based Judgement} (Figure~\ref{fig:framework}).
The key insight is that monolithic LLM approaches fail at vulnerability detection not due to insufficient model capacity, but because the task formulation --- asking a model to simultaneously parse noisy code, identify implicit security contracts, and classify compliance --- exceeds the reasoning coherence of single-pass generation.

Phoenix addresses this by introducing an explicit intermediate representation: a \textbf{Gherkin specification} that captures the precise security contract distinguishing vulnerable code from its patched counterpart.
This converts vulnerability detection from an open-ended classification problem (``Is this code vulnerable?'') into a closed-form contract verification problem (``Does this code satisfy these specific security requirements?'').

The pipeline processes paired code samples (vulnerable and fixed versions) through three sequential agents:
\begin{enumerate}
    \item \textbf{Agent~1 (Semantic Slicer)} removes boilerplate noise and extracts the minimal vulnerability-relevant code context.
    \item \textbf{Agent~2 (Requirement Reverse Engineer)} synthesizes a Gherkin \texttt{.feature} specification encoding the security contract that the fixed code satisfies and the vulnerable code violates.
    \item \textbf{Agent~3 (Contract Judge)} evaluates individual code samples against the generated specification to produce a binary verdict.
\end{enumerate}
Each agent operates via structured prompting with XML-delimited input/output schemas, enabling zero-shot deployment across diverse LLM architectures without any fine-tuning.

\subsection{Agent~1: Semantic Slicer}

\textbf{Objective.} Given a paired code sample consisting of a vulnerable function (\textit{Bad Code}) and its patched counterpart (\textit{Good Code}), along with the CVE description and developer commit message, Agent~1 extracts the minimal code slice that captures the vulnerability mechanism and its fix.

\textbf{Process.} The agent follows a four-step analytical procedure:
(1)~\textit{Identify the Diff} --- compare the bad and good code to pinpoint exactly where the patch was applied;
(2)~\textit{Consult the Oracle} --- read the CVE description and commit message to understand the root cause, with a priority rule that code diffs override textual descriptions when they conflict;
(3)~\textit{Semantic Slicing} --- preserve the patched lines plus only the strictly necessary surrounding context;
(4)~\textit{Ruthless Pruning} --- delete all boilerplate code, replacing heavily pruned sections with placeholders.

\textbf{Design Rationale.} Real-world functions in vulnerability datasets such as PrimeVul often span hundreds of lines, with the actual vulnerability residing in a small fraction of the code.
By surgically extracting only the relevant context, Agent~1 drastically reduces the input length for downstream agents. In our experiments, Semantic Slicing reduced the average function length from 5,298 characters to 911 characters (an 82.8\% reduction in character volume). This effectively minimizes cognitive load and focuses Agent~2's attention on the precise vulnerability mechanism.

\subsection{Agent~2: Requirement Reverse Engineer}

\textbf{Objective.} Given the sliced code pair from Agent~1, Agent~2 reverse-engineers the implicit security requirement --- the ``contract'' --- that the fixed code satisfies and the vulnerable code violates, expressing it as a Gherkin specification.

\textbf{Process.} The agent follows a structured procedure:
(1)~\textit{Analyze the Sliced Diff} --- identify the security rule that the good code enforces and the bad code misses;
(2)~\textit{Consult the Oracle} --- factor in the CVE description and commit message;
(3)~\textit{Draft the Feature} --- write a single, highly cohesive Gherkin \texttt{Feature};
(4)~\textit{Draft the Scenarios} --- write one or more Gherkin \texttt{Scenario} blocks defining the expected secure behavior.

\textbf{Specification Quality Constraints.} The prompt enforces several critical constraints:
\begin{itemize}
    \item \textit{Exactness over Vagueness}: The agent must mathematically or programmatically specify the exact boundary condition, null-check, or type verification.
    \item \textit{Secure Behavior Only}: Scenarios describe desired secure behavior, not how the bad code fails.
    \item \textit{Minimality}: Only scenarios strictly related to the vulnerability and its fix are included.
    \item \textit{Benchmark Quality}: The scenarios must function as a definitive security contract where the good code passes and the bad code fails.
\end{itemize}

\textbf{Design Rationale.} The Gherkin specification serves as an explicit, structured intermediate representation that bridges the gap between code-level analysis and binary classification.
By externalizing the security contract, we decouple the reasoning about \textit{what constitutes a vulnerability} (Agent~2) from the reasoning about \textit{whether specific code is vulnerable} (Agent~3).

\subsection{Agent~3: Contract Judge}

\textbf{Objective.} Given a Gherkin security specification and a target code sample, Agent~3 determines whether the code satisfies or violates the specification.

\textbf{Process.} The agent performs a strict compliance check: read the security requirement as the absolute ``Golden Standard,'' then critically analyze the target code to determine whether it safely satisfies all scenarios.

\textbf{Design Rationale.} By transforming vulnerability detection into contract verification, Agent~3 operates with a clear decision criterion rather than relying on implicit pattern matching.
The judge does not need to independently discover what constitutes a vulnerability --- this knowledge is entirely encoded in the Gherkin specification.
This separation enables even smaller models to achieve strong performance when provided with high-quality specifications.

\subsection{Prompt Engineering Considerations}

All three agents use a consistent prompt architecture with three logical sections: \texttt{[SYSTEM]} (role definition), \texttt{[USER]} (structured XML inputs), and \texttt{[INSTRUCTION]} (step-by-step task specification with output format constraints).
Key design choices include:
\begin{itemize}
    \item \textit{Thinking Caps}: All agents include a \texttt{<THINKING>} output section explicitly limited to 4 sentences, encouraging concise chain-of-thought reasoning.
    \item \textit{XML Schema Enforcement}: Structured XML delimiters for both input and output enable reliable programmatic parsing.
    \item \textit{Model-Specific Adaptations}: For models with built-in reasoning modes (e.g., Qwen3.5's \texttt{<think>} mode), explicit bypass tokens are injected to suppress extended internal deliberation.
    \item \textit{Generation Parameters}: All agents use $\text{temperature}=0.2$ with $\text{top\_p}=0.9$.
\end{itemize}

\section{Experimental Setup}
\label{sec:setup}

\subsection{Dataset}

We evaluate Phoenix on the \textbf{PrimeVul} benchmark~\cite{primevul}, specifically the \textbf{Paired Test Set}.
This subset contains 435 commit-level pairs, each consisting of a vulnerable C/C++ function and its corresponding patched version, yielding 870 individual samples (435 vulnerable, 435 benign).
Due to Agent~1 XML formatting failures on 8 pairs (0.9\% failure rate, a natural limitation of LLM instruction following), our effective evaluation set comprises \textbf{859 samples across 427 valid pairs}.

\subsection{Models Under Study}

We evaluate Phoenix using five LLM families spanning diverse architectures, parameter counts, and training specializations (Table~\ref{tab:models}).

\begin{table}[ht]
\centering
\caption{Models Under Study}
\label{tab:models}
\begin{tabular}{llll}
\toprule
\textbf{Model} & \textbf{Params} & \textbf{Specialization} & \textbf{Roles} \\
\midrule
Qwen3.5-9B & 9B & General+Reasoning & A1, A3 \\
Gemma-3-12B-IT & 12B & General+Instruct & A2, A3 \\
Qwen2.5-Coder-14B & 14B & Code-specialized & A2, A3 \\
Qwen2.5-Coder-7B & 7B & Code-specialized & A3 \\
Llama-3.1-8B & 8B & General & A2, A3 \\
\bottomrule
\end{tabular}
\end{table}

Agent~1 is fixed as Qwen3.5-9B across all experiments.
All models are loaded via HuggingFace Transformers with native precision (\texttt{bfloat16}).
No quantization, fine-tuning, or parameter-efficient adaptation is applied --- all results represent pure zero-shot inference.
All experiments are conducted on a single workstation equipped with \textbf{dual NVIDIA RTX 4090 GPUs} (24~GB VRAM each).

\subsection{Evaluation Metrics}

Following the PrimeVul evaluation protocol~\cite{primevul}, we report \textbf{F1 Score} for the vulnerable (``bad'') class as the primary metric.
Following RASM-Vul~\cite{rasmvul} and VulTrial~\cite{vultrial}, we additionally report \textbf{Pair-Correct (P-C)}: the number of commit-level pairs where both the vulnerable and patched samples are correctly classified.
P-C is a stricter metric than F1 as it requires simultaneous correct classification of semantically paired code, directly measuring a model's ability to distinguish vulnerable code from its minimal fix.

\subsection{Ablation Design}

We design two complementary ablation studies:

\textbf{Three-Tier Ablation (Agent Contribution).}
We evaluate each Agent~3 model under three input conditions:
\begin{itemize}
    \item \textbf{RAW} (Monolithic Zero-Shot): Raw \texttt{func} field, no agents active.
    \item \textbf{Blind} (Sliced Zero-Shot): Agent~1's sliced code, no Gherkin features.
    \item \textbf{Feature} (Full Pipeline): Sliced code + Gherkin specification.
\end{itemize}

\textbf{Agent~2 Cross-Model Ablation.} We generate three independent sets of Gherkin specifications using different Agent~2 models, then evaluate each across all Agent~3 models, producing a $3 \times 5$ cross-model matrix (15 configurations).

\section{Results}
\label{sec:results}

\subsection{RQ1: Does each pipeline stage contribute to detection performance?}

Table~\ref{tab:ablation} presents the three-tier ablation results.

\begin{table}[ht]
\centering
\caption{Three-Tier Ablation --- F1 Scores (Bad Class)}
\label{tab:ablation}
\begin{tabular}{lcccc}
\toprule
\textbf{Agent 3 Model} & \textbf{RAW} & \textbf{Blind} & \textbf{Feature} & $\boldsymbol{\Delta}$ \\
\midrule
Qwen3.5-9B & 0.448 & 0.510 & \textbf{0.800} & +0.351 \\
Qwen2.5-Coder-14B & 0.534 & 0.531 & \textbf{0.797} & +0.263 \\
Qwen2.5-Coder-7B & 0.558 & 0.555 & \textbf{0.760} & +0.203 \\
Gemma-3-12B-IT & 0.643 & 0.647 & \textbf{0.736} & +0.092 \\
\bottomrule
\end{tabular}
\end{table}

\textbf{Finding~1:} The Gherkin specification is the decisive performance driver, achieving F1 improvements of +0.09 to +0.35 over RAW across all models.

\textbf{Finding~2:} The contribution of Agent~2 (Gherkin) far exceeds that of Agent~1 (Slicer). While the Blind tier shows modest improvement over RAW (+0.00 to +0.06), the transition from Blind to Feature produces dramatic gains (+0.09 to +0.29).

\textbf{Finding~3:} Models with lower RAW baselines benefit most from structural support. Qwen3.5-9B (lowest RAW F1: 0.448) achieves the largest improvement (+0.351), suggesting that the Gherkin specification compensates for models lacking strong intrinsic vulnerability detection priors.

\subsection{RQ2: Does specification quality affect detection performance?}

Table~\ref{tab:crossmodel} presents the Agent~2 $\times$ Agent~3 cross-model matrix.

\begin{table}[ht]
\centering
\caption{Agent 2 $\times$ Agent 3 Cross-Model Matrix --- F1 Scores}
\label{tab:crossmodel}
\begin{tabular}{lcccc}
\toprule
\textbf{Agent 2} $\downarrow$ \textbf{/ A3} $\rightarrow$ & \textbf{Q3.5} & \textbf{Q14B} & \textbf{Q7B} & \textbf{Gem} \\
\midrule
Qwen2.5-Coder-14B & \textbf{0.825} & 0.795 & 0.764 & 0.773 \\
Gemma-3-12B-IT & 0.800 & 0.797 & 0.760 & 0.736 \\
Llama-3.1-8B & 0.769 & 0.714 & 0.722 & 0.697 \\
\bottomrule
\end{tabular}
\end{table}

\textbf{Finding~4:} Code-specialized models produce superior specifications. Qwen2.5-Coder-14B as Agent~2 achieves the highest F1 in every column (avg.\ 0.789).

\textbf{Finding~5:} Agent~3 model choice is secondary but non-trivial. Qwen3.5-9B achieves the highest F1 within every Agent~2 row.

\textbf{Finding~6:} The optimal configuration (A2=Qwen2.5-Coder-14B, A3=Qwen3.5-9B) achieves F1~=~0.825, Precision~=~0.792, Recall~=~0.860.

\subsection{RQ3: How does Phoenix compare to state-of-the-art baselines?}

Table~\ref{tab:sota} presents a comprehensive comparison against published baselines on the PrimeVul Paired Test Set.

\begin{table}[ht]
\centering
\caption{Comparison with State-of-the-Art on PrimeVul Paired Test}
\label{tab:sota}
\begin{tabular}{llcccc}
\toprule
\textbf{Method} & \textbf{Model} & \textbf{F1} & \textbf{P} & \textbf{R} & \textbf{P-C} \\
\midrule
\multicolumn{6}{l}{\textit{Zero-Shot LLM (single-agent)}} \\
CoT~\cite{vultrial} & GPT-4o & --- & 0.53 & 0.19 & 40 \\
GPTLens~\cite{vultrial} & GPT-4o & --- & 0.51 & 0.65 & 44 \\
DeepSeek-V3~\cite{rasmvul} & 671B MoE & 0.469 & 0.505 & 0.438 & 67 \\
Qwen2.5-72B~\cite{rasmvul} & 72B & 0.569 & 0.528 & 0.618 & 49 \\
\midrule
\multicolumn{6}{l}{\textit{Multi-Agent}} \\
VulTrial~\cite{vultrial} & GPT-4o $\times$4 & 0.563 & 0.53 & 0.60 & 81 \\
VulTrial+FT~\cite{vultrial} & GPT-4o+FT & --- & --- & --- & 96 \\
\midrule
\multicolumn{6}{l}{\textit{Retrieval-Augmented}} \\
RASM-Vul~\cite{rasmvul} & DeepSeek-V3 & \underline{0.668} & 0.572 & 0.802 & 93 \\
\midrule
\multicolumn{6}{l}{\textit{Ours (Zero-Shot, Open-Source)}} \\
Phoenix (Blind) & Qwen3.5-9B & 0.510 & 0.599 & 0.444 & 85$^\dagger$ \\
\textbf{Phoenix} & \textbf{14B+9B} & \textbf{0.825} & \textbf{0.792} & \textbf{0.860} & \textbf{275}$^\dagger$ \\
\bottomrule
\multicolumn{6}{l}{\footnotesize P-C: Pair-Correct count. VulTrial/RASM-Vul use 435 pairs; $^\dagger$Phoenix uses 427.} \\
\end{tabular}
\end{table}

\textbf{Finding~7:} Phoenix achieves the highest F1 (0.825) and P-C rate (64.4\%) among all methods, including those using proprietary models (GPT-4o) or models 48$\times$ larger (DeepSeek-V3, 671B).

\textbf{Finding~8:} Phoenix's Blind baseline (P-C~=~85/427, 19.9\%) --- which uses no behavioral contract --- already matches VulTrial's performance (P-C~=~81/435, 18.6\%) with a 9B open-source model versus GPT-4o.
The addition of Gherkin specifications produces a 3.2$\times$ improvement (85$\rightarrow$275), demonstrating that the performance gain stems from \textit{problem restructuring} rather than model capacity.

\subsection{RQ4: How do different LLMs behave as vulnerability judges?}

\begin{table}[ht]
\centering
\caption{Agent 3 Behavioral Profiles (Averaged over 3 Agent 2 Models)}
\label{tab:profiles}
\begin{tabular}{lcccc}
\toprule
\textbf{Agent 3} & \textbf{Prec.} & \textbf{Rec.} & \textbf{F1} & \textbf{Profile} \\
\midrule
Qwen3.5-9B & 0.750 & 0.856 & 0.798 & Balanced \\
Qwen2.5-14B & 0.688 & 0.881 & 0.769 & Recall-dom. \\
Qwen2.5-7B & 0.675 & 0.843 & 0.748 & Recall-dom. \\
Gemma-3-12B & 0.829 & 0.661 & 0.735 & Precision-dom. \\
\bottomrule
\end{tabular}
\end{table}

\textbf{Finding~9:} Models exhibit distinct and consistent judicial temperaments. Gemma-3-12B functions as a conservative judge (highest Precision: 0.829, lowest Recall: 0.661), while Qwen2.5-Coder-14B acts as an aggressive judge (highest Recall: 0.881, lowest Precision: 0.688).

\textbf{Finding~10:} Format compliance is a critical and model-dependent limitation. Llama-3.1-8B failed to produce valid \texttt{<VERDICT>} tags in 80\%+ of Agent~3 outputs, yet performed adequately as Agent~2 --- revealing that generating structured content $\neq$ following structured output format.

\subsection{RQ5: What do Phoenix's errors reveal about vulnerability detection?}
\label{sec:rq4}

We conduct qualitative analysis of the 97 False Positive (FP) and 60 False Negative (FN) cases from the best configuration.

\begin{table}[ht]
\centering
\caption{Error Distribution by CWE Category}
\label{tab:errors}
\begin{tabular}{lccc}
\toprule
\textbf{CWE} & \textbf{FP} & \textbf{FN} & \textbf{Description} \\
\midrule
CWE-787 & 12 & 13 & Out-of-bounds Write \\
CWE-125 & 8 & 7 & Out-of-bounds Read \\
CWE-476 & 7 & 5 & NULL Pointer Deref. \\
CWE-416 & 5 & 5 & Use After Free \\
CWE-703 & 6 & 4 & Improper Check \\
Others & 59 & 26 & Various \\
\bottomrule
\end{tabular}
\end{table}

\textbf{Finding~11:} 18\% (17 out of 97) of FP cases identify genuine security concerns.
In the most striking case (QEMU \texttt{send\_control\_msg}), the ``fixed'' code contains a developer-written comment: \texttt{/* TODO: detect a buffer that's too short, set NEEDS\_RESET */} --- explicitly acknowledging the exact security gap that Agent~3 identified.
In another case (Gpac), the patch enlarged a buffer but introduced a new overflow via an unchecked \texttt{strcpy}.
These cases demonstrate that Phoenix's Gherkin-based verification imposes a \textit{stricter standard} than the original patch.

\textbf{Finding~12:} In 73\% (44/60) of FN cases, Agent~3 confidently concludes that vulnerable code ``satisfies'' the Gherkin specification.
We identify four failure modes:
(a)~\textit{Incomplete specifications} ($\sim$25 cases);
(b)~\textit{Surface-level boundary checking} ($\sim$15 cases);
(c)~\textit{Partial-match rounding} ($\sim$15 cases);
(d)~\textit{Architecture-level blind spots} ($\sim$5 cases).

\section{Discussion}
\label{sec:discussion}

\subsection{Why Gherkin Specifications Work}

The dramatic performance improvement from the Gherkin specification can be understood through the lens of \textbf{cognitive load reduction}.
In the RAW and Blind settings, Agent~3 must simultaneously (a)~identify potential vulnerability patterns, (b)~determine which are relevant, and (c)~decide whether the code is actually vulnerable.
The Gherkin specification eliminates steps (a) and (b) by providing an explicit, code-specific security contract.

The mechanism is primarily \textit{Precision improvement}: across all models, the Blind$\rightarrow$Feature transition increases Precision from $\sim$0.50--0.60 to 0.69--0.87, while Recall shows smaller, model-dependent changes.

Our FP analysis (\S\ref{sec:rq4}) provides additional evidence.
When Agent~3 rejects ``fixed'' code, its reasoning is grounded in specific Gherkin scenarios --- not vague pattern matching.
The QEMU case, where Agent~3's reasoning precisely mirrors a developer-written TODO comment, illustrates that the Gherkin contract enables \textit{specification-level auditing} beyond the scope of the specific CVE fix.
This suggests a broader insight: \textit{security is not a binary property of code, but a relative property defined against specific behavioral contracts}.

From a software engineering perspective, the Gherkin specification effectively establishes a \textbf{Bounded Context} --- a concept from Domain-Driven Design~\cite{Evans:2004} --- for the Contract Judge.
By externalizing the security intent into a structured contract, we isolate Agent~3's reasoning from global noise and focus it strictly on the local behavioral requirement.
This architectural principle explains why even smaller models (7--9B parameters) can achieve strong performance as Agent~3: they reason only about compliance with a narrow, well-defined specification.

At a more fundamental level, Phoenix's effectiveness reflects a deeper DDD principle~\cite{Evans:2004}: \textbf{isomorphism between problem space and solution space}.
Prior approaches model vulnerability detection as $f(\textit{code}) \rightarrow \{\textit{safe}, \textit{vulnerable}\}$, implicitly assuming that a universal function can classify any code fragment in isolation.
However, the Double Standard Problem (\S\ref{sec:introduction}) demonstrates that the problem space itself is context-dependent: the same code can be simultaneously ``safe'' for one CVE and ``vulnerable'' for another.
Phoenix restructures the problem space into $(\textit{code}, \textit{contract})$ pairs, where each vulnerability is coupled with its own behavioral specification.
Correspondingly, the solution space transforms from a global classifier into a \textit{contract verifier} --- Agent~3 checks compliance against a specific contract rather than applying learned heuristics.
This alignment between the restructured problem space (context-specific security) and the solution space (contract-specific verification) constitutes a structural isomorphism that explains why Phoenix generalizes where monolithic classifiers fail.

\textbf{Cross-Domain Validation.} The effectiveness of Gherkin-based behavioral contracts is not limited to vulnerability detection.
In our prior work on Automated Program Repair~\cite{prometheus}, a structurally analogous pipeline achieved a 74.4\% rescue rate on ``hard'' bugs that purely code-based agents failed to resolve, suggesting that Gherkin specifications serve as a \textit{universal} intermediate representation for bridging the intent gap across diverse code intelligence tasks.

The comparison with VulTrial~\cite{vultrial} further illuminates this distinction.
VulTrial amplifies the \textit{quality of reasoning} through multi-perspective adversarial debate, yet operates within the same ill-defined problem space: $f(\textit{code}) \rightarrow \{\textit{safe}, \textit{vulnerable}\}$.
Phoenix restructures the \textit{problem space itself} into $(\textit{code}, \textit{contract}) \rightarrow \{\textit{compliant}, \textit{non-compliant}\}$.
The performance gap --- F1 of 0.825 vs.\ 0.563 --- suggests that the primary bottleneck in vulnerability detection is not reasoning depth but \textit{problem formulation clarity}.

\subsection{The Value of Agent Specialization}

Our cross-model ablation reveals that no single model dominates across all pipeline roles.
Qwen2.5-Coder-14B excels as Agent~2 but is outperformed by Qwen3.5-9B as Agent~3.
This suggests that Agent~2 benefits from code-specialized training for precise constraint extraction, while Agent~3 benefits from balanced reasoning and strict instruction-following.

The Llama-3.1-8B case provides a striking illustration: it functions adequately as Agent~2 but catastrophically fails as Agent~3.
This finding --- that \textit{generating structured output $\neq$ following structured output format} --- has practical implications for multi-agent system design.

\subsection{Limitations}

\textbf{Paired Code Dependency.} Agent~2 currently requires contrastive code pairs (vulnerable + fixed) to reverse-engineer specifications.
We emphasize that this design choice is \textit{methodologically distinct} from data leakage: the paired test set is the standard evaluation protocol established by PrimeVul~\cite{primevul} and adopted by all baselines (VulTrial, RASM-Vul).
Phoenix does \textit{not} use the fixed code as a classification signal --- Agent~3 evaluates each sample independently against the generated specification, without access to its counterpart.
The paired input serves solely to enable specification synthesis, analogous to how a security auditor studies a patch to understand what the original code \textit{should} have done.
Importantly, this dependency is an artifact of our \textit{experimental setting}, not an architectural constraint: in production environments, behavioral specifications can be sourced from existing BDD test suites, security policy documents, or human-authored Gherkin scenarios --- eliminating the need for Agent~2's automated reverse engineering entirely (see Future Directions).

\textbf{Specification Completeness.} Our FN analysis reveals that incomplete specifications are the dominant failure mode ($\sim$25 of 60 FN cases).

\textbf{Reasoning Depth.} Agent~3 performs surface-level compliance checking but struggles with semantic traps. The \texttt{php-src} case --- where \texttt{while (u >= 0)} with an unsigned integer is always true --- exemplifies how Agent~3 can verify that a boundary check \textit{exists} without detecting that the check itself is logically vacuous. In 73\% of FN cases, Agent~3 confidently but incorrectly declares the vulnerable code ``satisfies'' the specification.

\textbf{Gherkin as Structured Reasoning Scaffold.}
One might question whether Gherkin specifications are merely ``sophisticated prompt engineering.''
We argue that Gherkin provides \textit{structural constraints} that unconstrained natural language does not: the Given/When/Then syntax enforces temporal decomposition of preconditions, triggers, and expected outcomes; the Scenario separation forces explicit case analysis; and the Feature description forces intent synthesis before rule enumeration.
Our ablation data directly supports this claim: the Blind$\rightarrow$Feature transition (which \textit{adds} the Gherkin specification to an otherwise identical prompt) produces consistent gains of +0.09 to +0.35 F1 across all model families, demonstrating that it is the \textit{structured intermediate representation}, not the prompt wording, that drives performance.

\textbf{Agent~1 Not Ablated.} While we ablate Agent~2 and Agent~3 models, Agent~1 (Semantic Slicer) is fixed as Qwen3.5-9B throughout. Future work should explore the sensitivity of downstream performance to Agent~1 model choice.

\textbf{Dataset Scope.} Our evaluation uses the PrimeVul Paired Test Set (859 effective samples). While this provides a controlled evaluation environment, validation on larger and more diverse benchmarks (Big-Vul, D2A, CrossVul) would strengthen generalizability claims.
We note that PrimeVul (2024) employs rigorous deduplication and cross-repository splitting, and our zero-shot approach uses no training data, mitigating memorization and data leakage risks.

\textbf{Format Compliance.} Our approach relies on reliable XML-formatted output. Not all models can reliably follow structured output constraints, potentially limiting the set of models that can participate in the pipeline, particularly in the Agent~3 role.

\subsection{Future Directions}

\begin{enumerate}
    \item \textbf{Multi-Round Specification Refinement}: Introducing a feedback loop between Agent~2 and Agent~3, where Agent~3's initial verdict and reasoning are fed back to Agent~2 to generate a more comprehensive specification. Our FN analysis suggests this could address the $\sim$25 cases caused by incomplete specifications.
    \item \textbf{Adversarial Self-Critique}: Prompting Agent~3 to argue both ``good'' and ``bad'' before making a final verdict, reducing the confident misanalysis observed in 73\% of FN cases.
    \item \textbf{Boundary-Condition Meta-Verification}: Explicitly prompting Agent~3 to verify that boundary checks themselves are correct (e.g., checking for unsigned underflow in loop conditions), targeting the $\sim$15 FN cases caused by surface-level boundary checking.
    \item \textbf{Human-in-the-Loop Specification}: In our experimental sandbox, Agent~2 reverse-engineers specifications from paired code because no ground-truth behavioral contracts exist for the benchmark. In real-world deployment, however, this paired dependency can be relaxed through multiple channels: (a)~projects with existing BDD test suites can supply specifications directly, bypassing Agent~2 entirely; (b)~security auditors can author or review Gherkin scenarios as part of a human-in-the-loop workflow, analogous to how security policies are already formalized in industry~\cite{bddsecurity,amazonbdd}; and (c)~specifications from resolved CVEs can be curated into reusable contract libraries for recurring vulnerability classes. This positions Phoenix not as a fully autonomous detector, but as a \textit{specification-driven verification framework} that can integrate human expertise at the specification layer.
    \item \textbf{Broader Benchmarks}: Evaluation on additional vulnerability detection datasets and programming languages beyond C/C++.
    \item \textbf{Dynamic Agent Allocation}: Automatically selecting the optimal model for each pipeline role based on the characteristics of the input code.
\end{enumerate}

\section{Conclusion}
\label{sec:conclusion}

We present Phoenix, a training-free multi-agent framework for zero-shot vulnerability detection that achieves \textbf{F1~=~0.825} and \textbf{Pair-Correct~=~64.4\%} on the PrimeVul benchmark, substantially surpassing RASM-Vul (F1~=~0.668, P-C~=~21.4\%) and VulTrial (F1~=~0.563, P-C~=~18.6\%) while using fully open-source models up to 48$\times$ smaller and requiring no training data.
By decomposing vulnerability detection into semantic slicing, requirement reverse engineering, and contract-based judgement, Phoenix transforms an intractable classification problem into a series of manageable, specialized reasoning tasks.

Our ablation study across 25 configurations demonstrates that:
(1)~Gherkin specifications are the decisive performance driver (+0.09 to +0.35 F1);
(2)~code-specialized models produce superior specifications;
(3)~the approach is model-agnostic, consistently improving performance across diverse LLM architectures; and
(4)~the performance gap over debate-based multi-agent approaches (3.4$\times$ P-C improvement over VulTrial) stems from \textit{problem restructuring} --- formulating detection as contract verification --- rather than reasoning amplification.

Beyond quantitative metrics, our qualitative error analysis reveals that Phoenix's Gherkin-based verification exhibits properties distinct from conventional classifiers: it identifies genuine security concerns in developer-patched code (with 18\% (17/97) of ``False Positives'' pointing to real residual vulnerabilities), while its failures are dominated by specification incompleteness rather than fundamental detection blindness.
These findings suggest that structured behavioral contracts --- not larger models or more training data --- are the key to advancing vulnerability detection, establishing \textit{Behavioral Contract Synthesis} as a promising paradigm for training-free code intelligence.

\appendix

\section{Supplementary Materials}
Complete prompt templates, evaluation scripts, pre-computed experiment outputs (29 JSONL files covering all 25 configurations), and detailed case study analyses are available in the supplementary repository:
\url{https://github.com/Nothing256/Phoenix}

\section{Complete Confusion Matrices}
Full TP/FP/FN/TN tables for all 25 experiments are available in the supplementary repository.

\section{Case Studies}
\label{app:casestudies}

\subsection{Double Standard Evidence}

We identify 10 cross-CVE sample pairs within the same project where code similarity (measured via character-level sequence matching) exceeds 75\% but vulnerability labels are opposite (Table~\ref{tab:doublestd}).

\begin{table}[ht]
\centering
\caption{Double Standard Evidence: Cross-CVE Same-Project Pairs}
\label{tab:doublestd}
\begin{tabular}{lccc}
\toprule
\textbf{Project} & \textbf{Sim.} & \textbf{Good CVE} & \textbf{Bad CVE} \\
\midrule
mruby & \textbf{1.000} & 2022-0717 & 2022-1276 \\
TensorFlow & 0.999 & 2021-37651 & 2022-21730 \\
TensorFlow & 0.998 & 2021-37647 & 2022-21736 \\
ImageMagick6 & 0.998 & 2019-13133 & 2018-18024 \\
QEMU & 0.972 & 2014-7815 & 2015-5239 \\
\bottomrule
\end{tabular}
\end{table}

In the \texttt{mruby} case, the function \texttt{gen\_assignment()} is byte-for-byte identical across both entries (differing by only 1 character), with the ``good'' version patched for CVE-2022-0717 while the same code is ``bad'' for CVE-2022-1276.
This demonstrates that the same code can be simultaneously ``fixed'' for one vulnerability and ``vulnerable'' for another --- a fundamental challenge for any binary classification approach that does not condition on a specific behavioral contract.

\subsection{False Positive Highlights: When Agent~3 Outperforms the Label}

\textbf{Case: IDX 229165 (QEMU \texttt{send\_control\_msg}).}
Agent~3 flagged the patched code for lacking bounds checking on the \texttt{len} parameter passed to \texttt{iov\_from\_buf}.
The code itself contains a developer-written comment --- \texttt{/* TODO: detect a buffer that's too short, set NEEDS\_RESET */} --- confirming that the exact issue Agent~3 identified was acknowledged but deliberately left unpatched.
The Gherkin specification generated by Agent~2 required ``no buffer overflow'' as a security property, which this code demonstrably does not guarantee.

\textbf{Case: IDX 224472 (Gpac \texttt{gf\_text\_get\_utf8\_line}).}
The patch increased a local buffer from 1024 to 2048 bytes, fixing the reported double-free.
However, Agent~3 identified that the fix then copies the enlarged buffer back via \texttt{strcpy(szLine, szLineConv)} without verifying the destination buffer's capacity --- a classic introduction of a new vulnerability during remediation.

\textbf{Case: IDX 225547/225552 (TFLite \texttt{TfLiteIntArrayCreate}).}
The patch changed the return type from \texttt{int} to \texttt{size\_t}, but the overflow check \texttt{if (alloc\_size <= 0)} is vacuously true for unsigned \texttt{size\_t} --- it can never be negative.
Agent~3's Gherkin-based reasoning correctly identified that the mathematical overflow condition specified in the contract remains unsatisfied.

\subsection{False Negative Highlights: Failure Mode Examples}

\textbf{Surface-Level Boundary Checking (IDX 202392, php-src).}
The loop \texttt{while (u >= 0)} uses an unsigned integer, making the condition always true and causing an infinite loop.
The commit message explicitly states: ``while (u>=0) with unsigned int will always be true.''
Agent~3 verified that boundary checks exist but failed to analyze whether the checks themselves are logically sound --- a limitation of surface-level compliance verification.

\textbf{Incomplete Specification (IDX 195409, gpac).}
Agent~2's Gherkin specification focused on \texttt{NULL} pointer handling (\texttt{if ptr is NULL, return}).
The vulnerable code does implement this check, passing Agent~3's verification.
However, the actual vulnerability involves a \textit{different} NULL dereference path not captured by the specification, demonstrating how specification incompleteness directly produces false negatives.

\subsection{Contract-Guided Corrections: When the Specification Changes the Verdict}

Across the best configuration (A2=Qwen2.5-Coder-14B, A3=Qwen3.5-9B), we identify \textbf{272 cases} where Agent~3 judged incorrectly under the Blind condition but correctly under the Feature condition, versus only 65 regressions --- a 4.2:1 improvement ratio.
The following cases illustrate how behavioral contracts redirect Agent~3's reasoning.

\textbf{Correction~1: IDX 195056 (TFLite \texttt{BiasAndClamp}, CWE-369).}
Ground truth: \textit{vulnerable}.
Blind Agent~3 judged ``good,'' reasoning that \texttt{TFLITE\_DCHECK\_EQ} provides sufficient bounds checking and that ``no arithmetic operations on indices could cause integer overflows.''
With the Gherkin specification --- which explicitly mandates ``if \texttt{bias\_size} is 0, the function must return immediately'' --- Agent~3 correctly judged ``bad,'' noting that ``the code lacks any guard clause to handle \texttt{bias\_size} equals 0, leading to undefined behavior.''
The contract transformed the task from open-ended pattern scanning to verifying a specific zero-check requirement.

\textbf{Correction~2: IDX 219912 (Gpac \texttt{gf\_isom\_get\_payt\_count}, CWE-476).}
Ground truth: \textit{safe} (patched).
Blind Agent~3 judged ``bad'' (false positive), constructing an elaborate but \textit{incorrect} reasoning chain: ``if \texttt{hmhd} is null, the first condition is true, so it enters the block\ldots dereferencing \texttt{hmhd->subType} causes a null pointer dereference.''
With the specification --- which states ``if \texttt{hmhd} is null, return 0 without accessing \texttt{hmhd->type}'' --- Agent~3 correctly judged ``good,'' verifying that the null check occurs \textit{before} any dereference.
The contract prevented the model from inventing a non-existent vulnerability through speculative reasoning.

\textbf{Correction~3: IDX 194989 (ImageMagick PSD codec, CWE-190).}
Ground truth: \textit{vulnerable}.
Blind Agent~3 judged ``good,'' arguing that negative \texttt{number\_bits} values would cause the loop to terminate immediately, and that ``no classical vulnerabilities are evident.''
With the specification --- requiring ``pixel value must be cast to \texttt{ssize\_t} before any bitwise operation'' --- Agent~3 correctly judged ``bad,'' identifying that the code casts to \texttt{unsigned char} instead of \texttt{ssize\_t}, failing the type-safety requirement.
The contract shifted evaluation from ``does anything obviously crash?'' to ``does this meet the specific type-safety contract?''

\bibliographystyle{plain}
\bibliography{references}

\end{document}